\documentclass[prb,preprint]{revtex4}
\usepackage{amsfonts}
\usepackage{amsmath}
\usepackage{amssymb}
\usepackage{epsfig}
\usepackage{graphicx}%
\providecommand{\U}[1]{\protect\rule{.1in}{.1in}}
\newtheorem{theorem}{Theorem}
\newtheorem{acknowledgement}[theorem]{Acknowledgement}

\begin{document}
\title{{\Huge On Time and Space Double-slit Experiments}}
\author{M. Bauer}
\email{bauer@fisica.unam.mx}
\affiliation{Instituto de F\'{\i}sica, Universidad Nacional Aut\'{o}noma de M\'{e}xico,
M\'{e}xico DF, MEXICO}
\keywords{quantum interference; quantum transients; wave particle duality}
\pacs{PACS number}

\begin{abstract}
Time double-slit experiments have been achieved and presented as complementary
to spatial double-slit experiments, providing further confirmation of the
wave-particle duality. Numerical solutions of the free particle time dependent
Schr\"{o}dinger equation have been presented as explanation of the
experimental results. To be considered as exhibiting "interference in time"
has been objected to on the basis that the standard non relativistic quantum
theory does not have the property of coherence in time. In this note the
theoretical and experimental results are derived in a schematic but analytic
solution of the TDSE with appropiate initial boundary conditions. The time
evolution at a fixed position is shown to exhibit an oscillating transient
behavior. The particular boundary conditions are justified by the experimental
setups that actually result in having only a single electron at any given time
in the double-slit arrangement; and consequently achieve the construction of
double peak single electron wave packets, whose spreading gives rise to the
time behavior noted. The progressive complementarity of "which-path"
("which-time") information and "space interference" ("oscillating time
transient") pattern build up is also exhibited.

\end{abstract}
\maketitle

\section{\textbf{Introduction}}

In recent years, time electron double-slit experiments \cite{Lindner,Wollen}
have been achieved and presented as complementary to electron spatial
double-slit experiments \cite{Tonomura}, providing a further confirmation of
the wave-particle duality. A very important feature is that the experimental
conditions allow asserting that only one electron is present at any time. The
fringe patterns follow from the progressive accumulation of single particle events.

Interference from a double-slit allowed Thomas Young to demonstrate the wave
nature of light over two centuries ago. \cite{Young} However the explanation
by Einstein of the photoelectric effect in 1905, on the basis of Planck's
energy quantum hypothesis, followed by the Compton light scattering experiment
in 1923, brought forward evidence of a corpuscular behavior of light. On the
other hand, the daring assumption in 1924 by Louis de Broglie to conversely
associate a wave to matter was corroborated in 1927 by the Davisson-Germer
experiment of the diffraction of electrons by crystals. \cite{Beiser} Since
then the wave-particle duality in nature and the interpretation of quantum
mechanics have been the subject of extensive discussions and research.

The particle two-slit arrangement figured prominently as a thought experiment
in the exchange between Bohr and Einstein in the Solvay meetings of 1927 and
1930 \cite{Bohr} on the complementarity of the wave and particle aspects. As
expressed by Bohr, an attempt to detect through which slit the particle goes
excludes the development of an interference pattern. Einstein, on the other
hand, tried to override such assertion. As stated by Feynman, the double-slit
set up \textquotedblleft\ldots\ has in it the heart of quantum mechanics. In
reality it contains the only mystery\textquotedblright.\cite{Feynman}

The two-slit arrangement is included in most quantum mechanics textbooks to
illustrate the consequence of \ the de Broglie hypothesis.\cite{Feynman,
Gasio} Its experimental realization, however, had to wait more than 50 years
but has already involved electrons \cite{Tonomura, Fabioni}, neutrons
\cite{Gahler}, atoms \cite{atoms} and molecules.\cite{Nairz} Most striking is
the one carried out by Akira Tonomura and coworkers \cite{Tonomura} where the
buildup of the fringe pattern is achieved by the accumulation of time spaced
successive single electron impacts. The development of cavity quantum
electrodynamics (CQED) has been recently applied to carry out more
sophisticated \textquotedblleft which path\textquotedblright%
\ experiments.\cite{Haroche}

The development of double-slit experiments in the time domain brings out a new
facet to the wave-particle duality (\cite{Lindner,Wollen} and references
therein). The role of the slits is played by rapidly successive time windows
of very short duration. Also to be noted is that the question of diffraction
in time had been raised a long time before\cite{Moshinsky} and confirmed
experimentally only recently.\cite{ref} This is also included below, as in all
these cases, either an analytical solution \cite{Moshinsky} or numerical
integrations\cite{Lindner,Wollen} of the time dependent Schr\"{o}dinger
equation (TDSE) have been presented as the explanation of the results.

Notwithstanding, an objection has been raised to the claim of "interference in
time" on the basis that the non relativistic quantum theory does not have the
property of coherence in time.\cite{Horwitz} Indeed it is pointed out that
introducing two distinct packets into the beam of an experiment at two
different times would yield by construction a mixed state, for which no
interference would take place. An additional argument is that this would
require time to be an additional observable with a spectrum derivable from a
self-adjoint operator\cite{Horwitz}, as expansion of a state vector in such a
basis would provide superposition of different times. The existence of such an
operator is however a long standing problem in quantum
mechanics.\cite{Muga,Bauer,Hilgevoord} With respect to the diffraction in
time\cite{Moshinsky}, it is noted that the sudden lifting of a shutter does
result in damped transient type oscillations that can be interpreted as
fringes in time.

In this note it is shown, in a schematic analytic way, that both space and
time double-slit experiments, as well as the time diffraction, can indeed be
described by the TDSE for free particle motion with appropriate initial
boundary conditions in each case. The derivations account both for the
analytic structure and the numerical and experimental results. The time
dependence of the space density at a fixed point exhibits in all cases an
oscillatory transient type behaviour.

Finally, it is also shown that this formulation is adapted to exhibit the
progressive complementarity of \textquotedblleft which path\textquotedblright%
\ (\textquotedblleft which time\textquotedblright) information and
\textquotedblleft space interference\textquotedblright\ (\textquotedblleft
oscillating time transient\textquotedblright) pattern build up, as has already
been shown experimentally (\cite{Afshar,Mittel,Bach} and references therein).

\section{The free particle TDSE}

The time evolution of the state vector in the free particle case is given by:%

\begin{equation}
\left\vert \Psi(t)\right\rangle =e^{-i\hat{H}t/\hbar}\left\vert \Psi
(0)\right\rangle =e^{-i\boldsymbol{\hat{p}}^{2}t/2m\hbar}\left\vert
\Psi(0)\right\rangle
\end{equation}

Introducing the space and momentum representations one has:%
\begin{gather}
\left\vert \Psi(t)\right\rangle =%
{\displaystyle\int}
d\boldsymbol{r}e^{-i\boldsymbol{\hat{p}}^{2}t/2m\hbar}\left\vert
\boldsymbol{r}\right\rangle \left\langle \boldsymbol{r\mid}\Psi
(0)\right\rangle \nonumber\\
=(1/2\pi\hbar)^{3/2}%
{\displaystyle\int}
d\boldsymbol{p}\left\vert \boldsymbol{p}\right\rangle e^{-i\boldsymbol{\hat
{p}}^{2}t/2m\hbar}%
{\displaystyle\int}
d\boldsymbol{r}e^{-i\boldsymbol{p}\cdot\boldsymbol{r}/\hbar}\Psi
(\boldsymbol{r};0)
\end{gather}

This gives the state vector at time $t$ in terms of the initial space wave
function. It then follows that the wave functions in momentum space is:%

\begin{align}
\Phi(\boldsymbol{p};t)  &  =\left\langle \boldsymbol{p}\mid\Psi
(t)\right\rangle \nonumber \\
&  =(1/2\pi\hbar)^{3/2}e^{-ip^{2}t/2m\hbar}%
{\displaystyle\int}
d\boldsymbol{r}\text{ }e^{-i\boldsymbol{p}\cdot\boldsymbol{r}/\hbar}%
\Psi(\boldsymbol{r};0)
\end{align}
Its Fourier transform gives the wave function in configuration space
\cite{Gasio}, namely:
\begin{gather}
\Psi(\boldsymbol{r};t)=\left\langle \boldsymbol{r}\mid\Psi(t)\right\rangle
=\int d\boldsymbol{p}\left\langle \boldsymbol{r}\mid\boldsymbol{p}%
\right\rangle \left\langle \boldsymbol{p}\mid\Psi(t)\right\rangle \nonumber\\
=(m/2\pi\hbar t)^{3/2}%
{\displaystyle\int}
d\boldsymbol{r}^{\prime}e^{-im(\boldsymbol{r}^{\prime}-\boldsymbol{r}%
)^{2}/2\hbar t}\Psi(\boldsymbol{r}^{\prime};0)
\end{gather}

\bigskip

\subsection{\textbf{ Space double-slit}}

The initial condition is taken as:%
\begin{equation}
\Psi(\boldsymbol{r};0)=\text{ }\delta(x)\text{ }[\delta(y-a/2)+e^{-i\varphi
}\text{ }\delta(y+a/2)]\text{ }e^{ip_{0}z/\hbar}%
\end{equation}
corresponding to motion with initial momentum $p_{0}$ in the $z$ direction and
two point slits in the $y$ direction separated by a distance $a$. Dirac delta
functions are used for simplicity. Although mathematically they will spread
instantaneouly over all space, it will be seen that they do not modify the
essential results that narrow Gaussians properly normalizaed would yield. A
phase difference\textit{\ }$\varphi$ between the slits is introduced for
generality; it would appear in a Aharonov-Bohm set up.

Inserting (5) into (3), one easily obtains:%

\begin{equation}
|\Phi(\boldsymbol{p};t)|^{2}\text{ }\propto\text{ }\cos^{2}\left[
\frac{(p_{y}a/\hbar)-\varphi}{2}\right]
\end{equation}
\bigskip This shows that the particle acquires momentum in the $y$ direction
with alternating maxima and minima. Maxima are found at the values%
\begin{equation}
\lbrack(p_{y}a/\hbar)-\varphi]/2=\pi n,\text{\ \ \ }n=0,\pm1,\pm2,...
\end{equation}
corresponding to angles $\theta_{n}$ with the $z$ axis such that%
\begin{align}
\sin\theta_{n}  &  =(p_{y}/p_{0})=(2n\pi+\varphi)(\hbar/ap_{0}) \nonumber \\
&  =(n+\varphi/2\pi)\lambda_{B}/a
\end{align}
Here $\lambda_{B}$ = $h/p_{0}$ is the de Broglie wavelength of the incoming
particle. When $\varphi$ $=0$, there is a peak in the forward direction,
$\theta_{0}$ $=0$, with neighboring peaks at angles $\sin^{-1}(\pm\lambda
_{B}/a)$ . When $\varphi$ $\neq$ $0$, the interference pattern is shifted by
an angle $\theta=$ $\sin^{-1}[(\varphi/2\pi)\lambda_{B}/a]$. These are well
known textbook results. \cite{Gasio} It is also easily seen that the
interference pattern disappears if one of the holes is shut, as proven in
Section III.

The time dependent wave function (4) yields now:%
\begin{equation}
|\Psi(\boldsymbol{r};t)|^{2}\propto(2m\hbar/t)^{3}\cos^{2}[(amy/2\hbar
t)+\varphi/2]
\end{equation}
where $y$ is the transverse coordinate where the interference pattern is
observed. For $y\neq0$, besides the damping factor $t^{-3}$, the intensity
oscillates very rapidly for small $t$ and slows down to almost constant for
$t$ large. One\ has therefore a transient-type pattern\cite{Gaston} with a
time dependent period $\Xi(t)$ that increases with time as $\Xi(t)=(2\pi
\hbar/amy)t^{2}.$ With $\varphi=0$ and $\theta<<1$, one has as initial value
$\Xi_{n}(T_{0})=T_{0}/n$ at the secondary peaks of the interference pattern at
a distance $Z$ corresponding to $Y_{n}=Z\tan\theta_{n}\approx Z\theta_{n}$ and
classical arrival time $T_{0}=Z/v_{0}=mZ/p_{0}$.\ 

Textbook presentations of the space double-slit case never include analysis of
the time evolution as given here. Perhaps an experiment like that of Tonomura
and coworkers could provide a confirmation.\cite{Tonomura} Although the
electron source is continuous (field-emission electron microscope), the
current density is so low as to have only one electron in the system at any
time (the time of flight of a 50 keV electron to cover a source detector
distance of 1.5 m. is about 11 ns, while the current density is 3000 electrons
per second). This allows to record individual arrivals at the scintillation
detector, which are then projected successively into the television screen as
time progresses. The grouping into fringes shows that the accumulation rate of
events (number of electrons per unit time) exhibits a space-dependence. But,
as noted above from Eq.9, the accumulation rate also has an oscillating
time-dependence. It seems it would be a question of following the accumulation
rate of successive recordings of one specific detector if the arrival times
are stored in the computer.

\subsection{\textbf{Diffraction in time}}

Diffraction in time concerns the much earlier identification of the occurrence
of transient effects in a dynamical description of resonance
scattering.\cite{Moshinsky,Gaston} It describes the effect of the opening of a
single shutter at a certain time. It is included here for completeness since
it is another TDSE development that can be treated in the same schematic way.%


The initial condition is now taken as:%
\begin{equation}
\Psi(\boldsymbol{r};0)=\text{ }\delta(x)\text{ }\delta(y)\text{ }%
\theta(-z)\text{ }e^{-ip_{0}z/\hbar}%
\end{equation}
where $\theta(z)$ is the Heavyside step functions. This corresponds to a plane
wave with momentum $p_{0}$ in the positive $z$ direction confined to negative
\ $z$ \ until a point orifice at the origin is opened at time $t=0.$ Inserting
(10) into (4) yields for $t\geq0$:%
\begin{align}
\Psi(\boldsymbol{r};t)  &  =%
\frac12
e^{-i3\pi/2}(m/2\pi\hbar t)e^{[-im(x^{2}+y^{2})/2\hbar t]}\nonumber\\
&  \times e^{-imz^{2}/2\hbar t}e^{Y_{0}^{2}}\operatorname{erf}\text{c}(Y_{0})
\end{align}
where $\operatorname{erf}$c$(Y_{0})$ is the complementary error function
($\operatorname{erf}$c=1-$\operatorname{erf}$),$\mathit{\ }Y_{0}=e^{-i\pi
/4}(2\hbar t/m)^{-1/2}[z-v_{0}t]$ and $v_{0}=p_{0}/m$. The $z$-dependence
coincides exactly with the one dimensional wave function of Eqs.3a, 3b, 3c of
Ref.12. From these equations, the ratio of the transient to the stationary
current density (no shutter) at a point $Z$ ahead of the shutter corresponding
to a classical arrival time $T=Z/v_{0}$, is calculated and exhibited in Fig.1
to be compared with Fig.3 of Ref.12. The signal begins to build up at $T$ and
exhibits a decaying transient behavior afterwards.

\subsection{Time double-slit}

Consider a superposition at a time $t$ of two state vectors evolving under the
same conditions from different initial times $t_{1}$and $t_{2}$ with
$t_{1}<t_{2}.$Then, as the unitary evolution operator $U(t,t^{\prime})$
satisfies the relation $U(t,t^{\prime})=U(t,t^{\prime\prime})U(t^{\prime
\prime},t^{\prime})$:%
\begin{align*}
\left\vert \Psi(t)\right\rangle  &  =\text{ }U\left(  t,t_{1}\right)
\left\vert \Phi(t_{1})\right\rangle \text{ }+\text{ }U\left(  t,t_{2}\right)
\left\vert \Theta(t_{2})\right\rangle \text{ }\\
&  =U\left(  t,t_{2}\right)  U\left(  t_{2},t_{1}\right)  \left\vert
\Phi(t_{1})\right\rangle +\text{ }U\left(  t,t_{2}\right)  \left\vert
\Theta(t_{2})\right\rangle \text{ }%
\end{align*}
Thus:%
\[
\left\vert \Psi(t)\right\rangle =U\left(  t,t_{2}\right)  \{\left\vert
\Phi(t_{2})\right\rangle +\left\vert \Theta(t_{2})\right\rangle \}
\]
Then $\left\vert \Psi(t)\right\rangle $ is given by the evolution from the
superposition at the same instant$\ \ t_{2}$ . On this basis, with
$\Phi=\Theta=\Psi,$ the initial condition $(t_{2}=0)$ for the time double-slit
is taken as:%

\begin{align}
\Psi(\boldsymbol{r};0)  &  =\text{ }\delta(x)\text{ }\delta(y)\text{ } \nonumber \\
&  \times\lbrack\delta(z-a/2)+e^{-i\varphi}\text{ }\delta(z+a/2)]\text{
}e^{ip_{0}z/\hbar}%
\end{align}
It consists of two pulses moving in the $z$ direction with velocity\textit{\ }%
$v=p_{0}/m$ , separated by a distance $a=(p_{0}/m)\tau$ where $\tau$ is the
time delay between pulses\textit{.} A phase shift is introduced that may be
related to the pulse creation mechanism. \cite{Lindner,Wollen} Inserting (12)
into (3) one obtains:%
\begin{align}
|\Phi(\boldsymbol{p};t)|^{2}\text{ }  &  \propto\text{ }\cos^{2}\left[
\frac{(p_{z}-p_{0})a/\hbar-\varphi}{2}\right] \nonumber \\
&  =\cos^{2}\left[  \frac{(p_{z}-p_{0})(p_{0}\tau/m\hbar)-\varphi}{2}\right]
\end{align}
with alternating maxima and minima. The peaks occur at momenta $p_{z}\ =p_{n}$
such that%
\begin{equation}
p_{n}=p_{0}+(m\hbar/p_{0}\tau)[2\pi n+\varphi],\text{\ \ \ \ }n=0,\pm
1,\pm2,....
\end{equation}

In terms of energy one has peaks at%
\[
E_{n}=p_{n}^{2}/2m=E_{0}+\frac{\hbar}{\tau}[2\pi n+\varphi]+\frac{\hbar^{2}%
}{4E_{0}\tau^{2}}[2\pi n+\varphi]^{2}%
\]
where $E_{0}=p_{0}^{2}/2m$\textit{\ }. Thus, neglecting the second term, the
separation $\delta E$ between consecutive peaks is given by $\ 2\pi\hbar
/\tau=h$\textit{/}$\tau$, as exhibited in Fig.1b of Ref.2 . More precisely
$\delta E$ $\geq h/\tau$ or $\tau\times\delta E\geq h$. There is thus a
complementary relation of the time delay with the energy interference pattern.

The time evolution of the space wave function (4) yields:%
\begin{gather}
|\Psi(\boldsymbol{r};t)|^{2}\text{ }\propto\left(  \frac{2m\pi\hbar}%
{t}\right)  ^{3}\cos^{2}[\frac{p_{0}\tau}{2m\hbar}(p_{0}-mz/t)+\varphi
/2]\nonumber\\
=\left(  \frac{2m\pi\hbar}{t}\right)  ^{3}\cos^{2}[\frac{E_{0}\tau}{\hbar
}(1-\sqrt{mc^{2}/2E_{0}}(z/ct))+\varphi/2]
\end{gather}
This exhibits the spread of the original wave packets (in this case infinite
because of the delta pulses) that gives rise to a fringe type pattern, as well
as the overall damping with time to conserve probability. At a fixed position
$z$ it consists of a damped oscillation (the $t^{-3}$ factor) with a period
increasing with time, namely $\Xi(t)=(\pi\hbar/E_{0}\tau)(p_{0}/mz)t^{2}.$ It
oscillates very rapidly for small $t$\ and flattens down to almost constant
for $t$ large. It is a transient response.\cite{Gaston} For finite width
peaks, the transient response would begin to be detected at a point $Z$ at the
classical time the pulse gets there, namely $T_{z}=Z/v_{0}=mZ/p_{0}$ , as in
the case of the shutter and in Ref.2. The period of oscillation would then
have the value $\Xi(T_{z})=(\pi\hbar/E_{0}\tau)T_{z}$ , increasing thereafter
from this value. This behavior\ is exhibited in Fig.2, where the position $Z$
and classical arrival time $T$ correspond to those of Fig.1e of Ref.2.

For fixed $t$, the space probability density (15) exhibits maxima at%

\begin{align}
(p_{0}\tau/2m\hbar)(p_{0}-mz/t)  &  =(p_{0}^{2}\tau/2m\hbar)(1-mz/p_{0}t) \nonumber \\
&  =2\pi n-\varphi/2
\end{align}
or, equivalently at%
\begin{equation}
\lbrack(E_{0}\tau/\hbar)(1-\sqrt{\frac{mc^{2}}{2E_{0}}}\frac{z}{ct})]=2\pi
n-\varphi/2,\ \ \ n=0,\pm1,\pm2,...
\end{equation}
where $E_{0}=p_{0}^{2}/2m$ , $v_{0}=p_{0}/m$ \textit{and }$c$ the velocity of light.

The above expressions can now be compared with the experimental and calculated
results of Refs.2 and 3, where measurements are made in the energy domain and
related to the time delay between wave packet peaks and their time evolution.
As stated in Ref.2, the cosine function oscillates, at a given time $t$ and
for a given $\varphi$, with variations of:

- momentum $p_{0}$ (and thus energy $E_{0}=p_{0}^{2}/2m$), for $\tau$ and $z$ fixed;

- delay time $\tau$, for $p_{0}$ and $z$ fixed;

- distance $z$ from the origin, for $p_{0}$ and $\tau$ fixed.

Furthermore, there is a displacement in the $z$ direction with time given by%
\begin{equation}
z=v_{0}t=(p_{0}/m)t=(2E_{0}/m)^{1/2}t
\end{equation}
reflecting the conservation of momentum in the free particle motion.

The dependence on the incident energy in Eq.17 allows comparison with the
measured energy spectra in Fig.2 of Ref.1, as well as the exchange of maxima
and minima with a $\pi/2$\ change in phase. As shown following Eq.14 above,
the separation between consecutive maxima is $\delta E\approx h/\tau.,$that
for $\tau$ = 2 fs yields $\delta E\approx$\ 2 eV. In an interval of 14 eV, one
then expects seven peaks, in agreement with the experimental results in Fig.2
and the numerical simulation in Fig.3 of that paper.

For an electron energy $E_{0}$= 0.3 eV and a time delay between pulses $\tau
$=120 fs, Eq.18 yields $z$ =113 nm at $t$ =350 fs, $z$ = 293 nm at $t$ = 900
fs and $z$ =1626 nm at $t$ =5000 fs, values that roughly correspond to the
calculated wave fronts in Figs.1(c,d,e) of Ref 2. Also for $\tau$ = 96 fs,
$h$/$\tau$ = 43 meV, in agreement with the calculated energy peak separation
in Fig.1b and the experimental one in Fig.3d of this reference.

It is easily shown (Section III) that the oscillation disappears if one of the
temporal slits in (12) is supressed, in agreement with the results of \ Ref.1
when only one temporal slit is generated.

\section{Complementarity in wave-particle duality}

Complementary weights can be assigned to each slit in the space double-slit
case by taking the initial wave function to be:%
\begin{align}
\Psi(\boldsymbol{r};0)  &  =\text{ }\delta(x)\text{ }[\alpha\delta
(y-a/2)+(1-\alpha)e^{-i\varphi}\text{ }\delta(y+a/2)]\text{ } \nonumber \\
&  \times\delta(z)\text{ }e^{ip_{0}z/\hbar}%
\end{align}
where $\alpha$ varies between $0$ and $1$. The values $0$ and $1$ correspond
to having only one slit open, while intermediate values correspond to
partially blocking one while opening the other. Substitution in Eq.3 yields:%
\begin{gather}
|\Phi(\boldsymbol{p};t)|^{2}\text{ }\propto\text{\ }\alpha^{2}+(1-\alpha
)^{2}+2\alpha(1-\alpha)\cos[(p_{y}a/\hbar)-\varphi]\nonumber\\
=(2\alpha-1)^{2}+4\alpha(1-\alpha)\cos^{2}[\frac{(p_{y}a/\hbar)-\varphi}{2}]
\end{gather}
which clearly shows the fading of the interference pattern as $\alpha$
approaches either $0$ or $1$. For $\alpha=1/2,$ Eq.6 is recovered, which
corresponds to maximum visibility of the interference pattern. This
progressive complementarity has been recently confirmed experimentally by
placing a movable mask in front of a double slit to control the transmissions
through the individual slits.\cite{Bach}

An entirely similar result is obtained when applied to the time double-slit
case, where the extreme $\alpha$-values $0$ and $1$correspond to having only a
single pulse and consequently no oscillating transient pattern.\cite{Wollen}

\section{Conclusions}

It has been shown in an analytic schematic way that the free particle TDSE
does give rise to the calculated and observed space and time double-slit
results, as well as the diffraction in time, when initial boundary conditions
appropriate to the experimental or theoretical set up are considered.
Moshinsky's "diffraction in time" actually corresponds to the fact that the
transient behavior generated by the shutter opening has the appearance of a
Fresnel pattern.\cite{Moshinsky,Horwitz,Gaston} It is seen here that the time
evolution of the wave function in the time and space double-slit experiments
exhibits also an oscillatory transient response.

The reservation raised in Ref.16 is related to whether states corresponding to
different times can give rise to a coherent superposition, and thus generate
cross terms in the probability density in some representation, to be
interpreted as interference terms; or alternatively, they can only give rise
to a mixture. It indeed can be shown that two states distinguished in some way
(e.g, by a particular quantum number) do not allow a superposition but only a
mixed state.\cite{K} As an example, the construction of a beam inserting at
one time electrons \textit{with spin up} and at another time electrons
\textit{with spin down} is presented in Ref.16\ . Although being correct, it
does not apply here.

The success of the numerical solutions of the TDSE in Refs.1 and 2, as well as
of the analytic developments presented here, revolves on whether the
experimental setups\cite{Lindner,Wollen,Tonomura} actually result in having
only a single electron at any given time in the double-slit arrangement. In
the space double-slit experiment this is achieved by a very low flux ("the
average interval of successive electrons is 1.5 m. In addition, the length of
the electron wave packet is as short as $\sim$ 1 $\mu$%
m".\cite{Tonomura,Tonomura2} In the time double-slit experiments,
photoionization is induced by two time-delayed femtosecond laser pulses ("So
far the free interfering electrons are originating neither from double
ionization of one atom nor from single ionization of different
atoms"\cite{Wollen}) or by phase stabilized few-cycle laser pulses of
femtosecond duration that open one to two windows (slits) of attosecond
duration ("The temporal slits leading to electrons of given final momentum are
spaced by approximately the optical period"\cite{Lindner}). It is then claimed
that the wave packets thus generated have to be considered as one double
peaked free electron wave packet, as has been assumed in this paper. Indeed
one is considering a single wave packet amplitude that happens to have
initially two peaks that do not overlap; in the plane perpendicular to the
direction of motion in the case of the space double slit; in the direction of
motion in the case of the time double slit. These peaks spread in time and
will eventually overlap and cause the probability density to oscillate as a
function of time at any space location.This is supported by the results of the
experiments and constitutes an extraordinary technical achievement.

Finally, it is also shown (Section III) that the progressive closing of one of
the space slits (or time slits) results in the progressive disappearence of
the interference pattern as the "which-path" ("which-time") information is
affirmed. This is in agreement with experiments that have indeed revealed the
possibility of partial fringe visibility and partial which-path information
(\cite{Afshar,Mittel,Bach} and references therein). \ 

\begin{acknowledgement}
The author wishes to thank Profs. L.P. Horwitz , P.A. Mello and the unknown
reviewers for their critical but illuminating observations, as well as Profs.
M. Fortes and J. Flores for their assistance in the preparation and revision
of the manuscript.
\end{acknowledgement}

\bigskip

\textbf{Figure Captions}

\textbf{Figure 1:} Ratio of the transient to the stationary current at a point $Z=v_{0}T$, where $T$ is the classical arrival time.

\textbf{Figure 2:} Transient response at Z=1626 nm from the classical time of arrival T(5000 fs) to 2T ($E_{0}=0.3$ eV, $\tau=120$
fs)



\end{document}